\documentclass{article}

\usepackage{amsmath}
\usepackage{amssymb}
\usepackage{color}
\usepackage[linesnumbered,longend,ruled,vlined]{algorithm2e}
\usepackage{tikz}
\usepackage{booktabs}
\usetikzlibrary{arrows}
\usepackage{subcaption}
\usepackage{multicol}
\usepackage{changes}
\usepackage{rotating}
\usepackage{longtable}
\usepackage{url}

\title{Conditional Markov Chain Search for the Simple Plant Location Problem improves upper bounds on twelve K\"orkel-Ghosh instances}

\author{Daniel Karapetyan and Boris Goldengorin}

\begin{document}

\maketitle

\begin{abstract}
 We address a family of hard benchmark instances for the Simple Plant Location Problem (also known as the Uncapacitated Facility Location Problem).
 The recent attempt by Fischetti et al.~\cite{fischetti2017} to tackle the K\"orkel-Ghosh instances resulted in seven new optimal solutions and 22 improved upper bounds.
 We use automated generation of heuristics to obtain a new algorithm for the Simple Plant Location Problem.
 In our experiments, our new algorithm matched all the previous best known and optimal solutions, and further improved 12 upper bounds, all within shorter time budgets compared to the previous efforts.
 
 Our algorithm design process is split into two phases: (i) development of algorithmic components such as local search procedures and mutation operators, and (ii) composition of a metaheuristic from the available components.
 Phase (i) requires human expertise and often can be completed by implementing several simple domain-specific routines known from the literature.
 Phase (ii) is entirely automated by employing the Conditional Markov Chain Search (CMCS) framework.
 In CMCS, a metaheuristic is flexibly defined by a set of parameters, called configuration.
 Then the process of composition of a metaheuristic from the algorithmic components is reduced to an optimisation problem seeking the best performing CMCS configuration.
 
 We discuss the problem of comparing configurations, and propose a new efficient technique to select the best performing configuration from a large set.
 To employ this method, we restrict the original CMCS to a simple deterministic case that leaves us with a finite and manageable number of meaningful configurations.
 
%

\end{abstract}

\section{Introduction}

 The Simple Plant Location Problem (SPLP), also known as Uncapacitated Facility Location Problem, is a classical combinatorial optimisation problem~ Cornuejols et al.~\cite{cornuejols90} with many applications in quantitative logistics,  Daskin~\cite{daskin2013} and flexible manufacturing systems, Goldengorin et al.~\cite{goldengorin2013}. 
 The SPLP takes a set $I = \{1, 2, \ldots, m \}$ of sites in which plants can be located, a set $J = \{1, 2, \ldots, n \}$ of clients, each having a unit demand, a
vector $F = (f_i)$ of fixed costs for setting up plants at sites $i \in I$, and a matrix $C = [c_{i j}]$ of transportation costs from $i \in I$ to $j \in J$ as input. It computes a set $P^\star$,
$\emptyset \subset P^\star \subseteq I$, at which plants can be located so that the total cost of satisfying all client demands is minimal. 
 The costs involved in meeting the client demands include the fixed costs of setting up plants, and the transportation cost of supplying clients from the plants that are set up. 

Historical roots of the SPLP can be found in pioneering Weber's publication~\cite{weber1929}, and a modern formulation of the SPLP as a Mixed Integer Linear Programmming (MILP) problem can be read in Balinski~\cite{balinski1965}.

 A detailed introduction to this problem has appeared in Cornuejols et al.~\cite{cornuejols90}. 
 The SPLP forms the underlying model in several combinatorial problems, such as set covering, set partitioning, information retrieval, simplification of logical Boolean expressions, airline crew scheduling, vehicle despatching, and is a subproblem for various location analysis problems  (see Goldengorin et al.~\cite{goldengorin2003} and references within).

The SPLP is $\mathcal{NP}$-hard (Cornuejols et al.~\cite{cornuejols90}), and several exact and heuristic algorithms for solving it have been discussed in the literature. Most of the exact algorithms are based on a mathematical programming formulation of the SPLP (see for example, 
Cornuejols and Thizy~\cite{cornuejols1982}, Morris~\cite{morris1978}, and Schrage~\cite{schrage1975}). Polyhedral results for the SPLP polytope have been reported in Trubin~\cite{trubin1969}, Balas and Padberg~\cite{balas1972}, Cho et al.~\cite{cho1983a}, Cho et al.~\cite{cho1983b}, Farias~\cite{farias2001}, C{\'a}novas et al.~\cite{canovas2002}, and Galli et al.~\cite{galli2016}. 
 In theory, these results allow us to solve the SPLP by applying the simplex algorithm to the strong linear programming relaxation, with the additional stipulation that a pivot to a new extreme point is allowed only when this new extreme point is integral. However, efficient implementations of this pivot rule are not available. Beasley~\cite{beasley1993} reported computational experiments with Lagrangian heuristics for SPLP instances. K\"{o}rkel~\cite{korkel1989} proposed algorithms based on refinements to a dual-ascent heuristic procedure to solve the dual of a linear programming relaxation of the SPLP combined with the use of the complementary
slackness conditions to construct primal solutions Erlenkotter~\cite{erlenkotter1978}. Barahona and Chudak~\cite{barahona2000}, \cite{barahona2005} have reported optimal solutions to some SPLP instances with $m = n =3000$ and paid attention to computationally difficult SPLP instances with large fixed costs and several opened sites in an optimal solution and easy solvable SPLP instances with small fixed costs and almost all opened sites in an optimal solution. 

 Since the SPLP is NP-hard, an essential number of publications are devoted to approximation and heuristic algorithms (see e.g., Resende and Werneck~\cite{resende2006}). For example, Guha and Khuller~\cite{guha1999} have established a lower bound of 1.463 for the approximation factor, under some widely believed assumptions. 
 Another heuristic by Jain et al.~\cite{jain2003}, has a performance guarantee of only 1.61, but in computational experiments returns good quality SPLP solutions within 2\% of their optimality. 
 In practice, these heuristics tend to be much closer to optimality for non-pathological instances. 
 There is a long list of heuristics without any theoretically proven approximation ratio for the found feasible solutions which return high quality SPLP solutions. 
 Among them constructive and local search heuristics rooted from the pioneering work of Kuehn and Hamburger~\cite{kuehn1963}, and successfully continued by simulated annealing Alves and Almeida~\cite{alves1992} and Yigit et al.~\cite{yigit2004}, genetic algorithms Kratica et al.~\cite{kratica2001}, complete local search with memory Ghosh~\cite{ghosh2003}, and tabu search Michel and P. Van Hentenryck~\cite{michel2003} as well as Sun~\cite{sun2006}. 
 Dual-based methods, such as Erlenkotters~\cite{erlenkotter1978} dual ascent, Guignard's~\cite{guignard1988} Lagragean dual ascent, and the volume algorithm by Barahona and Chudak~\cite{barahona2000} have also shown promising results. 
 An experimental comparison of some state-of-the-art heuristics is presented by Hoefer~\cite{hoefer2003} with a recommendation that tabu search finds the highest quality heuristic solutions within reasonable CPU time.
 
Researchers found that many SPLP instance families are relatively easy to solve.
For example, one can solve all Beaslye SPLP benchmark instances just by two Khumawala  preprocessing rules combined with a few branchings on variables with the largest violation within a fraction of a second Goldengorin~\cite{goldengorinhco2013}.
Letchford and Miller~\cite{letchford2014} designed preprocessing rules which are effective for the SPLP  instances with facilities and clients located at points on the Euclidean plane.
In 2003, Ghosh~\cite{ghosh2003} proposed a class of computationally hard instances which are now known as K\"orkel-Ghosh (KG) instances since these instances are modified K\"orkel instances~\cite{korkel1989}. 
Fischetti et al.~\cite{fischetti2017} explain the computational intractability of the KG instances because they have a large number of near-optimal solutions, which makes it hard to identify variables that could not be in an optimal KG SPLP instance solution. Since on average at least 80\% of all sites should be closed in an optimal solution to the KG instances most of pre-processing approaches are not successful in their efforts to find high quality solutions to the KG instances.
 The KG instance library includes three classes of instances, namely, A, B and C.
 In class A, the fixed costs $f_i$ are drawn uniformly from $[100, 200]$, in class B -- from $[1000, 2000]$, and in class C -- from $[10000, 20000]$.
 The transportation costs $c_{ij}$ are always drawn uniformly from $[1000, 2000]$.
 Symmetric and asymmetric instances are included, where symmetric instances satisfy $c_{ij} = c_{ji}$.
 The KG library includes instances of size $m \times n = 250 \times 250$, $m \times n = 500 \times 500$ and $m \times n = 750 \times 750$.
 
 In the recent decade, many different heuristics (see e.g. Sun~\cite{sun2006}) as well as exact approaches by Beltran-Royo et al.~\cite{beltran2012}, Posta et al.~\cite{posta2014}, Fischetti et al.~\cite{fischetti2017} were applied to improve the best known upper bounds for the KG instances.
 A recent attack on the KG benchmark showed that the upper bounds for many of the instances can still be improved Fischetti et al.~\cite{fischetti2017} but this takes a significant computational effort. Fischetti et al.~\cite{fischetti2017} conclude that 50 KG instances 
still remain out of reach for existing exact methods. 
 Nevertheless, they have been able to improve the best known upper bounds for 22 KG instances solutions and matched the other 21 within 3,600 seconds. 
 After increasing the CPU time budget to 7,200 second they slightly improved their results by keeping 22 strictly improved and 1 more matched (now 22 matched) solutions. 
 For the remaining 6 instances (out of 50) their upper bounds are worse than the best known.

 The purpose of our paper is to present the next step in finding better solutions to the KG SPLP benchmark instances.
 While all the previous attempts to tackle SPLP were based on human-designed algorithms, we applied automated heuristic generation to produce an effective method for SPLP\@.
 
 The main idea behind automated heuristic generation is that (meta)heuristic design is a labour-intensive process in which an expert is required to use their skills and intuition about the domain to combine available components into an algorithm with complex behaviour.
 Automated generation of (meta)heuristics, also known as generating hyper-heuristics, is meant to make the design process cheaper and quicker, and avoid the subjective judgement of the expert that usually affects the algorithm architecture.
 The completely automated algorithm design is not yet available, however a recent approach called Conditional Markov Chain Search (CMCS) enables one to automatically compose a metaheuristic from a set of given domain-specific routines in Karapetyan et al.~\cite{Karapetyan2017}.
 
 The CMCS gives a flexible framework capable of describing a wide range of metaheuristics using a set of parameters.
 Each specific combination of parameter values is called a \emph{configuration}.
 In other words, a configuration is a specific composition of a metaheuristic fro the available domain-specific routines. 
 By selecting one of the top performing configurations, we generate an effective metaheuristic.
 
 We use the this approach to automatically design a simple yet effective metaheuristic for SPLP\@.
 An important contribution of the paper is a refined CMCS generation method.
 Observe that the problem of selecting the best performing configuration out of several candidates is not well-defined, mainly because there is unlikely to be a single configuration performing better than every other configuration in every test.
 We propose an approach to selecting the best CMCS configuration from the space of all feasible configurations.
 We further apply several rules to significantly reduce the space of CMCS configurations and use a brute-force-like algorithm to choose the best of them.
  
 While searching for the best performing CMCS configuration, we use a training dataset consisting of small instances, and use short running times.
 Nevertheless, the performance of our selected configuration scales well to the size of large KG instances.
 In particular, we show that our automatically generated metaheuristic clearly outperforms previous state-of-the-art heuristics including the most recent computational records in Fischetti et al.~\cite{fischetti2017}.
 Moreover, in our experiments it improved 12 (and matched 38 remaining) best known values among the 50 yet unsolved KG SPLP instances, and have not returned any worse solution for all previously solved 90 KG benchmark instances keeping the total CPU time budget not more than 1 second!

 The paper is structured as follows.
 The SPLP-specific parts of the algorithm, i.e.\
 the data structures and algorithmic components, are described in Section~\ref{sec:components}.
 The CMCS framework, the generation procedure and the best performing CMCS configurations are discussed in Section~\ref{sec:cmcs}.
 The computational results of applying the best performing CMCS configurations to the benchmark instances are reported in Section~\ref{sec:computational-results}.
 The concluding remarks and future work are discussed in Section~\ref{sec:conclusions}.

\section{SPLP Components}
\label{sec:components}

 In this section we describe the domain-specific components that will later be used within our CMCS configurations.
 All of these components are well-known from the SPLP literature, or are variations of standard algorithms.

\subsection{Data structures}
\label{sec:data-structure}

 Our data structure is based on the ideas previously proposed in the SPLP literature, see e.g.~\cite{Hansen1997}.
 We store the list of opened sites in two forms: a vector $y \in \{0,1\}^m$, where $y_i$ indicates whether site $i$ is opened, and a set of indices $P \subseteq I$ of opened sites.
 In addition, for each client $j \in J$, we store the closest opened site $p(j) \in P$ and the second closest opened site $q(j) \in P$.
 Thus, the objective value of a solution can be efficiently computed as
 \[
	\sum_{i \in P} f_i + \sum_{j \in J} c_{p(j),j} \,.
 \]
 In practice, we never need to compute the objective value as we store it in a variable $v$ and maintain its value while manipulating the solution.
 
 Our data structure requires that at least two sites are opened, which is a reasonable assumption for our test problems, and so we enforce this constraint in every component.
 If for some other problem instances such an assumption would be too strong, one can start the search with evaluating all the $m$ solutions containing exactly one opened site, which would take only $O(mn)$ time.
 
 We also pre-compute a matrix $\pi = [\pi(i,j)]$ of size $m \times n$, where $\pi(i,j)$ is the index of the $j$th closest site for the client $i$.
 In other words, $c_{\pi(i,1)} \le c_{\pi(i,2)} \le \cdots \le c_{\pi(i,m)}$.
 We will need the matrix $\pi$ for efficient exploration of a neighbourhood, see Section~\ref{sec:open-best}.
 
 This data structure allows efficient procedures for opening or closing a site.
 For details see Algorithms~\ref{alg:open} and \ref{alg:close}.
 The worst case time complexity of opening a site is $O(n)$ and of closing -- $O(n |P|)$.
 
\SetKwInOut{Input}{input}
\SetKwInOut{Output}{output}

\begin{algorithm}[htb]
	\caption{Opening a site}
    \label{alg:open}
    
    \Input {Site $i^* \in I \setminus P$ to be opened}
    
	\ForAll {$j \in J$}
    {
    	\If {$c_{i^*,j} < c_{p(j), j}$}
        {
            $v \gets v - c_{p(j), j} + c_{i^*, j}$\;
        	$q(j) \gets p(j)$\;
        	$p(j) \gets i^*$\;
        }
    }    
    
    $v \gets v + f_{i^*}$\;
    $P \gets P \cup \{ i^* \}$\;
    $y_{i^*} \gets 1$\;
\end{algorithm}

\begin{algorithm}[htb]
	\caption{Closing a site}
    \label{alg:close}
    
    \Input {Site $i^* \in P$ to be closed}
    
    $P \gets P \setminus \{ i^* \}$\;
    $y_{i^*} \gets 0$\;
	\ForAll {$j \in J$}
    {
    	\If {$p(j) = i^*$}
        {
            $v \gets v - c_{p(j), j} + c_{q(j), j}$\;
        	$p(j) \gets q(j)$\;
            $q(j) \gets \arg\min_{i \in P \setminus \{ p(j) \}} c_{i,j}$\;
        }
        \ElseIf {$q(j) = i^*$}
        {
            $q(j) \gets \arg\min_{i \in P \setminus \{ p(j) \}} c_{i,j}$\;
        }
    }    
    $v \gets v - f_{i^*}$\;
\end{algorithm}

\subsection{Open Random ($k$)}

 The first two components we discuss are mutation operators, i.e.\ components that make random changes to the solution, usually applied to escape a local minimum by worsening its quality.
 The `Open Random ($k$)' component opens $k$ randomly selected sites.
 More specifically, the component selects $k$ distinct sites, opening those of them that are not currently opened (we assume that the number of opened sites is relatively small and thus the probability of hitting a site that is already opened is relatively small).
 
 The time complexity of the `Open Random ($k$)' component is $O(kn)$.

\subsection{Close Random ($k$)}

 The `Close Random ($k$)' component is another mutation operator; it closes $k$ randomly selected sites.
 More specifically, the component selects $\min \{ k, |P| - 2 \}$ currently opened sites and closes them.
 The worst case time complexity of the `Close Random ($k$)' component is $O(kn |P|)$.

\subsection{Open Best}
\label{sec:open-best}

 The `Open Best' component is a local search procedure that opens a single site if that improves the solution.
 The cardinality of the corresponding neighbourhood is $m - |P|$, and the procedure chooses the best candidate.
 A naive implementation of the `Open Best' local search would take $O(mn)$ time.
 We reduce this to $O(m + \sum_{j \in J} p(j))$.
 (Observe that $p(j) \le m$ and hence $\sum_{j \in J} p(j) \le mn$, while in practice the sum is considerably smaller.)
 This is achieved by gradually building a vector $\delta_i$, $i \in I$, where $\delta_i$ is the change in the objective value if the site $i$ is to be opened.
 We initialise $\delta_i \gets f_i$.
 Then, for each client $j \in J$, we scan through the sites $i$ that are closer than $p(j)$, i.e.\ through the sites that, if opened, will improve the transportation cost for that client, and update $\delta_i \gets \delta_i + c_{i,j} - c_{p(j),j}$.
 This operation is implemented efficiently by utilising the precomputed $\pi$ matrix, see Section~\ref{sec:data-structure}.
 At the end, we choose $i^* = \arg\min_{i \in I} \delta_i$.
 If $\delta_{i^*} < 0$ then we open site $i^*$.
 Otherwise we leave the solution unchanged.

\subsection{Close Best}

 The `Close Best' component is a local search procedure that closes a single site if that improves the solution.
 The corresponding neighbourhood consists of $|P|$ solutions, and the procedure chooses the best out of them.

 A naive implementation of the `Close Best' local search would take $O(|P| n)$ time, whereas we reduce the exploration time to $O(m + n)$ time.
 We gradually build a vector $\delta_i$, $i \in I$, where $\delta_i$ is the change in the objective value caused by closing site $i$.
 Initially we set $\delta_i = -f_i$ for every $i \in P$.
 Then, for each client $j \in J$, we increase $\delta_{p(j)}$ by $c_{q(j), j} - c_{p(j), j}$.
 At the end, we choose $i^* = \arg\min_{i \in P} \delta_i$, and if $\delta_{i^*} < 0$ then we close site $i^*$.
 Otherwise we leave the solution unchanged.

\subsection{Exchange Best}
 
 The `Exchange Best' component is a local search procedure that closes one site and opens another one if that improves the solution.
 It chooses the best candidate out of the $(m - |P|) |P|$ solutions in the neighbourhood.
 
 A native implementation would take $O(mn |P|)$ time to explore the `Exchange Best' neighbourhood.
 By following the logic of the `Open Best' local search implementation and building a matrix $\delta$ of size $|P| \times n$, we reduce this to $O(mn)$.

\subsection{Exchange Half Fixed}

 Observe that the `Exchange Best' local search is relatively slow comparing to the other two local search procedures.
 We propose a local search that explores only a fraction of the `Exchange Best' neighbourhood but runs much quicker.
 Our `Exchange Half Fixed' local search randomly selects the site $i^* \in P$ to be closed, and then searches for the best site to be opened.
 This exploration takes $O(\sum_{j \in J} p(j) + m \gamma)$ time, where $\gamma$ is the number of clients $j$ for which $p(j) = i^*$.
 For a random instance, the expected value of $\gamma$ is $n / |P|$.

\section{Conditional Markov Chain Search}
\label{sec:cmcs}

\newcommand{\Mfail}{M^\text{fail}}
\newcommand{\Msucc}{M^\text{succ}}
\newcommand{\ComponentPool}{\mathcal{H}}

 Conditional Markov Chain Search (CMCS) was first introduced by Karapetyan et al.~\cite{Karapetyan2017} as a framework to enable automated generation of metaheuristics.
 It gives a flexible way of composing a metaheuristic from a set of domain-specific components, with the behaviour of the control mechanism defined by numerical parameters.
 
 Let $\ComponentPool$ be an ordered set of available domain-specific components, which we call a \emph{solution pool}.
 By component we mean a black box algorithm that takes the problem instance and a solution as an input and outputs a new (modified) solution.
 A components could implement, for example, a local search procedure or a random move (mutation).
 
 CMCS is a single-point metaheuristic.
 It applies the components to the current solution, one at a time, in a certain sequence.
 The output of the previous component is an input of the next component in the sequence.
 For example, if the sequence is 
 $$
 \ComponentPool_1, \ComponentPool_1,\ComponentPool_3, \ComponentPool_3, \ComponentPool_2, \ComponentPool_1,
 $$
 and the initial solution is $S_0$, the CMCS will proceed as follows:
 \begin{align*}
	S_1 \gets \ComponentPool_1(S_0),\\
	S_2 \gets \ComponentPool_1(S_1),\\
	S_3 \gets \ComponentPool_3(S_2),\\
	S_4 \gets \ComponentPool_3(S_3),\\
	S_5 \gets \ComponentPool_2(S_4),\\
	S_6 \gets \ComponentPool_1(S_5).
 \end{align*}
 CMCS saves the best solution found so far.
 Hence, at the end of iteration $i$, it stores two solutions: current solution $S_i$ and the best of $S_0$, $S_1$, \ldots, $S_i$.
 
 Each component modifies the solution according to its internal logic.
 The change may improve or worsen the solution quality; a component may also leave the solution intact.
 
 The components are stateless, and are independent (do not communicate with each other).
 A component may be randomised or be deterministic.
 
 Given a fixed set of components, the behaviour of CMCS is defined by the sequence in which the components are executed.
 This sequence is decided online.
 The decision of which component to execute in iteration $i$ is made at the end of iteration $i - 1$.
 In particular, this decision depends on two factors: (i) which component was executed at iteration $i - 1$, and (ii) whether $S_{i - 1}$ is better than $S_{i - 2}$.
 Hence, the sequence of components is a Markov chain, with the state consisting of the last executed component and a Boolean variable indicating whether the last executed component has improved the solution.
 
 The specific logic of the next component selection is called \emph{configuration}.
 While CMCS is a framework, a CMCS configuration is a fixed metaheuristic algorithm.
 Observe that a CMCS configuration can be completely defined by two transition matrices, $M^\text{succ}$ and $M^\text{fail}$, each of size $|\ComponentPool| \times |\ComponentPool|$.
 The matrix $\Msucc$ is used when the solution is improved by the last executed component, and $\Mfail$ is used otherwise (when either the objective value has not changed, or is has worsened).
 To keep the paper self-contained, we include a pseudo-code of CMCS in Algorithm~\ref{alg:cmcs}, a close copy from~\cite{Karapetyan2017}.
 
\begin{algorithm}[htb]
\Input{Components pool $\mathcal{H}$;}
\Input{Matrices $\Msucc$ and $\Mfail$ of size $|\ComponentPool| \times |\ComponentPool|$;}
\Input{Objective function $f(S)$ to be minimised;}
\Input{Instance data $\mathcal{I}$;}
\Input{Initial solution $S_0$;}
\Input{Termination time $\mathit{terminate\text{-}at}$;}

$S^* \gets S_0$\;
$f^* \gets f(S_0)$\;
$f_\text{prev} \gets f^*$\;
$h \gets 1$\;
$i \gets 1$\;
\While {$\mathit{now} < \mathit{terminate\text{-}at}$}
{
	$S_i \gets \ComponentPool_h(\mathcal{I}, S_{i-1})$\;
    $f_\text{cur} \gets f(S_i)$\;
    \If {$f_\text{cur} < f_\text{prev}$}
    {
	    $h \gets \mathit{RouletteWheel}(\Msucc_{h, 1}, \Msucc_{h, 2}, \ldots, \Msucc_{h, |\ComponentPool|})$\;
        
        \If {$f_\text{cur} < f^*$}
        {
        	$S^* \gets S_i$\;
            $f^* \gets f_\text{cur}$\;
        }
    }
	\Else
	{
	    $h \gets \mathit{RouletteWheel}(\Mfail_{h, 1}, \Mfail_{h, 2}, \ldots, \Mfail_{h, |\ComponentPool|})$\;
    }
    
    $f_\text{prev} \gets f_\text{cur}$\;
    $i \gets i + 1$\;
}

\Return {$S^*$}\;

\caption{Conditional Monte-Carlo Search}
\label{alg:cmcs}
\end{algorithm}

 CMCS does not have any acceptance criteria, i.e.\ it never backtracks any changes made by the components.
 (Backtracking can be implemented within a domain-specific component, e.g.\ inside a local search procedure, however once the component execution is finished, the change, in general, cannot be undone.)
 Therefore, the only source of the improvement pressure in CMCS is the improvement pressure generated by some of the components.
 As a result, it is necessary to include in the component pool at least one component that would be biased towards good solutions, such as a local search procedure.
 It is equally important to include at least one component capable of worsening the solution, such as a mutation operator, to escape local minima.
 
 Observe that CMCS is completely domain-independent as all the knowledge of the domain is incorporated in the components treated as black boxes.
 Hence, both the control mechanism and the configuration generation routines are domain-independent and reusable.
 Such a reusability is a long-standing goal in the area of optimisation algorithm design.

 We proceed by discussing in Section~\ref{sec:configuration-enumeration} how to restrict CMCS to leave only a finite manageable number of configurations, and also how to enumerate them, and then, in Section~\ref{sec:select-configuration}, we discuss how to choose the best of the available configurations.

\subsection{Deterministic CMCS}
\label{sec:configuration-enumeration}

 Recall that a CMCS configurations is specified by two matrices, $M^\text{succ}$ and $M^\text{fail}$.
 Each value in a transition matrix defines the probability of the corresponding transition, hence the space of configurations is continuous.
 Searching in this space is particularly hard due to the roughness of the landscape, typical in parameter tuning.
 However, as shown in~\cite{Karapetyan2017}, discretisation of the search space allows one to use brute force to optimise some special cases of CMCS\@.
  
 In this project, we restrict CMCS to the deterministic case, i.e.\ to the case where each row of $M^\text{succ}$ and $M^\text{fail}$ contains exactly one non-zero element.
 (Note that the resultant configuration is not necessarily a deterministic SPLP algorithm; it is only the transition mechanism that is deterministic.)
 This leaves us with $k^{2k}$ feasible configurations.

 Out of these configurations, some are equivalent.
 Consider the example in Figure~\ref{fig:equivalent-configurations}.
 As $\ComponentPool_3$ is unreachable in either of the two configurations, the last row of the matrices can be ignored (it does not affect the behaviour of the configurations).
 Then the two configurations, formally different, are equivalent.

\begin{figure}[htb]
	\centering
    \begin{subfigure}[t]{1\textwidth}
    $$
    \Msucc = \left[\begin{array}{rrrr}
    	& \ComponentPool_1 & \ComponentPool_2 & \ComponentPool_3 \\
        \ComponentPool_1 & 1 & 0 & 0 \\
        \ComponentPool_2 & 1 & 0 & 0 \\
        \ComponentPool_3 & 0 & 1 & 0 \\
    \end{array}\right]
    \qquad
    \Mfail = \left[\begin{array}{rrrr}
    	& \ComponentPool_1 & \ComponentPool_2 & \ComponentPool_3 \\
        \ComponentPool_1 & 0 & 1 & 0 \\
        \ComponentPool_2 & 0 & 1 & 0 \\
        \ComponentPool_3 & 1 & 0 & 0 \\
    \end{array}\right]
    $$
    \caption{Configuration A.}
    \end{subfigure}
    \hfill
    \begin{subfigure}[t]{1\textwidth}
    $$
    \Msucc = \left[\begin{array}{rrrr}
    	& \ComponentPool_1 & \ComponentPool_2 & \ComponentPool_3 \\
        \ComponentPool_1 & 1 & 0 & 0 \\
        \ComponentPool_2 & 1 & 0 & 0 \\
        \ComponentPool_3 & 0 & 0 & 1 \\
    \end{array}\right]
    \qquad
    \Mfail = \left[\begin{array}{rrrr}
    	& \ComponentPool_1 & \ComponentPool_2 & \ComponentPool_3 \\
        \ComponentPool_1 & 0 & 1 & 0 \\
        \ComponentPool_2 & 0 & 1 & 0 \\
        \ComponentPool_3 & 1 & 0 & 0 \\
    \end{array}\right]
    $$
    \caption{Configuration B.}
    \end{subfigure}
\caption{
	Configurations A and B are equivalent.
    Indeed, they are only different in the last row of $\Msucc$, i.e.\ in transition from $\ComponentPool_3$.
    However, $\ComponentPool_3$ is unreachable.
    Hence, the last row of either of the transition matrices does not affect the behaviour of the configurations.
}
\label{fig:equivalent-configurations}
\end{figure}
 
 To exclude such `duplicates', we follow a two-step procedure: 
\begin{enumerate}
	\item
    At first, we generate all non-empty subsets $H \subseteq \ComponentPool$.
    
    \item
    For each subset $H$, we generate all the configurations that use every component $h \in H$.
    By `use' we mean that there exists a non-zero probability of transition from any $h' \in H$ to $h$, perhaps within several iterations.
    This can be formalised using a directed graph $G = (H, E)$ with a node set $H$ and arc set $E$ which includes an arc $(h, h') \in E$ if and only if $\Msucc_{h,h'} + \Mfail_{h,h'} > 0$.
    We say that the configuration uses all the components $H$ if and only if graph $G$ is strongly connected. 
    One may note that some component $h$ may not be reachable from some other components, however be executed during the first iterations of the algorithm, for example if it is the entry point.
	We assume here that the effect of $h$ in such a case is likely to be negligible after a large number of iterations.
\end{enumerate}

 We can further eliminate some configurations by imposing several constraints:
\begin{itemize}
    \item
   	At least one of the components in $H$ needs to generate improvement pressure.
    In practice, this usually means that at least one of the components is a local search.
    
    \item
    At least one of the components in $H$ needs to be able to worsen the solution, as otherwise the search will quickly converge to a local minimum and stop there.
    In practice, this usually means that as least one of the components is a mutation.
    
    \item
    If component $H_h$ is a classic local search, i.e.\ it explores some deterministic neighbourhood and makes the move if and only if it improves the solution, then we can fix $\Mfail_{h,h} = 0$.
\end{itemize}

 We say that a configuration that satisfies all the above conditions is \emph{meaningful}.
 
 This still leaves us with a considerable number of meaningful configurations.
 For example, for a set of six components, where three of them are deterministic local search procedures and three are mutations, the number of meaningful configurations is approximately $3.4 \cdot 10^8$. 
 This is a significant improvement over the number of feasible configurations $k^{2k} \approx 2 \cdot 10^9$, but still impractical even for such a small number of components.
 Thus, we introduce an additional constraint; we only consider configurations for $|H| = \lambda$, where $\lambda$ is a parameter.
 Then we can ask a question of the form ``what is the best configuration composed of exactly $\lambda$ components'' or ``what is the best configuration composed of at most $\Lambda$ components''.
 
 The parameter $\lambda$ greatly reduces the number of configurations.
 The number of feasible $\lambda$-component configurations is 
\begin{equation}
	\label{eq:feasible}
	\binom{|\ComponentPool|}{\lambda} \cdot \lambda^{2\lambda} \text{ configurations.}
\end{equation}
 Given three local searches and ten mutation, there are only $2.1 \cdot 10^5$ three-component and $1.2 \cdot 10^3$ two-component feasible configurations.
 By using the conditions discussed above, we end up with $3.7 \cdot 10^4$ three-component and $1.8 \cdot 10^2$ two-component meaningful configurations (there are no meaningful configurations with one component, as we require that both local searches and mutations are included into $H$, see the above constraints).
 Compare this to the overall $9.2 \cdot 10^{28}$ feasible configurations.

 
 We recognise that the parameter $\lambda$ greatly restricts the complexity of the CMCS configurations, however this restriction allows us to include many components and let the CMCS generator, rather than a human expert, choose which component combinations are most efficient.

\subsection{CMCS Generator}
\label{sec:select-configuration}

\newcommand{\TrainingData}{\mathcal{T}}

 CMCS generator is a procedure that finds the best (or some very good) CMCS configuration.
 In this project, as we restrict the set of configurations to deterministic configurations, our CMCS generator aims at selecting the best of all the meaningful configurations.
 
 To evaluate a configuration, we use a training dataset $\TrainingData$.
 Each element of $\TrainingData$ is a triple $(\mathit{Inst}, S_0, t)$, where $\mathit{Inst}$ is the SPLP instance, $S_0$ is the initial solution, and $t$ is the time budget.
 Let $f(C, \mathit{Inst}, S_0, t)$ be the objective value of a solution obtained by solving instance $\mathit{Inst}$ with the initial solution $S_0$ and the time budget $t$ by the CMCS configuration $C$.
 Then we can interpret the problem of selecting the best configurations as a multi-dimensional optimisation problem, with the objective functions $f(C, \mathit{Inst}, S_0, t)$, $(\mathit{Inst}, S_0, t) \in \TrainingData$.
 With a large number of dimensions, one may assume that the majority of the configurations would be Pareto optimal.
 However, in practice many configurations demonstrate very poor solution quality and as a result are dominated by top ranked configurations.
 Hence, the Pareto domination approach is sufficient to filter out the majority of poorly performing configurations.
 
 The approach taken in Karapetyan et al.~\cite{Karapetyan2017} was to run all the tests for each configuration.
 Here we improve this by filtering out the least promising configurations after the first few tests.
 More specifically, if a configuration $C$ performs strictly worse than some other configuration $C^*$ in the first seven tests, we can use the sign test condition to conclude that $C^*$ is superior to $C$ with significance level 99\%.
 This heuristic approach will not allow us to select the best performing configuration but it will let us quickly focus on the most promising configurations.
 
 Selection of the best performing configuration out of the the most promising candidates requires multiple-criteria decision-making.
 The standard methods, such as the Analytic hierarchy process or ELECTRE, are designed to tackle problems with hard to quantify and compare attributes.
 In our problem, all the attributes (the $f(C, \mathit{Inst}, S_0, t)$ values) have equal weights, and are inherently easy to quantify.
 Thus, we use using the simple weighted sum model, with equal weights.
 
 To summarise, our CMCS generator performs in two stages:
\begin{enumerate}
	\item
	Form a set $\mathcal{C}$ of all meaningful configurations and run the first seven tests $\TrainingData_1, \TrainingData_2, \ldots, \TrainingData_7$ for each configuration $C \in \mathcal{C}$.
    Select non-dominated configurations and save them to $\mathcal{C}'$, i.e. 
    \begin{equation}
    \mathcal{C}' = \{ C \in \mathcal{C} :\; \not\exists C^* \in \mathcal{C} \text{ such that } C^* \text{ dominates } C \} \,.
    \end{equation}
    $\mathcal{C}'$ usually includes only a small fraction of all the meaningful configurations.
    
	\item
    Run the remaining tests for each configuration in $\mathcal{C}'$.
    As the scale of objective values may vary between tests, then, for each $(\mathit{Inst}, S^0, t) \in \TrainingData$, normalise $f(C, \mathit{Inst}, S^0, t)$ by scaling it to the $[0, 1]$ interval:
    $$
    f'(C, \mathit{Inst}, S^0, t) = \frac{f(C, \mathit{Inst}, S^0, t) - \min_{C' \in \mathcal{C}'} f(C', \mathit{Inst}, S^0, t)}{\max_{C' \in \mathcal{C}'} f(C', \mathit{Inst}, S^0, t) - \min_{C' \in \mathcal{C}'} f(C', \mathit{Inst}, S^0, t)} \,.
    $$
    Finally, select a configuration $C \in \mathcal{C}'$ that minimises 
    \begin{equation}
    	\sum_{(\mathit{Inst}, S^0, t) \in \TrainingData} f(C, \mathit{Inst}, S^0, t) \,.
    \end{equation}
    
\end{enumerate}

\section{Computational Results}
\label{sec:computational-results}

 We first describe in Section~\ref{sec:splp-cmcs-generation} our set up for the CMCS configuration generation, as well as the produced configurations.
 We then apply in Section~\ref{sec:experiments} these configurations to the KG instances to obtain upper bounds for the yet unsolved instances and compare our results to the state-of-the-art heuristics from the literature.
 We also check the performance of our CMCS configurations on already solved instances.
 
 All our algorithms were implemented in C\#, and the experiments were conducted on a Windows machine with two Intel Xeon E5-2690 v4 (2.6 GHz) CPUs and HyperThreading enabled.
 Our implementation of CMCS does not use concurrency.
 We used the multiple cores to run the experiments in parallel, but not more than one experiment per physical CPU core.

\subsection{CMCS configuration generation for SPLP}
\label{sec:splp-cmcs-generation} 
 
 In our experiments, we restricted the set of configurations to two- and three-component configurations, i.e.\ set $\Lambda = 3$.
 Our pool $\ComponentPool$ of components consisted of: 
 \begin{itemize}
 	\item
 	`Open Best', `Close Best', `Exchange Best' and `Exchange Half Fixed' local searches;
    
    \item
    `Open Random ($k$)' and `Close Random ($k$)' mutations for $k = 1, 2, 3, 4$.
 \end{itemize}
 
 The training data set included 200 tests $(\mathit{Inst}, S_0, t)$.
 The instances $\mathit{Inst}$ were generated using the KG instance generator, with the instance size $n = m$ selected uniformly at random between 300 and 400.
 The size and the type of the instance (`a', `b' or `c') was also drawn uniformly at random.
 The initial solution $S_0$ was generated by opening $r$ random sites, where $r \in [2, \lfloor 0.1 n \rfloor]$ was chosen uniformly at random.
 Note that an optimal solution to a KG instance is likely to have more than two but less than $\lfloor 0.1n \rfloor$ opened sites; hence, we exercise both situations when the number of opened sites needs to be increased and decreased.
 The time budget $t$ for each test was set to 0.5 sec.
 This time budget was selected to allow a CMCS configuration to run a sufficiently large number of iterations (about 50) to reveal its long term behaviour.
 
 Some of the data on CMCS generation is summarised in Table~\ref{tab:cmcs-generation}.
 One can see that the value of $\lambda$ significantly affects the number of meaningful configurations, as well as the number of configurations selected at the first stage of the generation.
 As a result, the wall time taken by the generator for $\lambda = 2$ was around four minutes whereas for $\lambda = 3$ it was more than twelve hours.
 This shows that the current generator is not well suited for more complex configurations or significantly larger component pools, however even this limited set of configurations yields good results, as we will show in our computational study.

\begin{table}[htb]
\begin{center}
\begin{tabular}{@{} lrr @{}}
\toprule

	& $\lambda = 2$
    & $\lambda = 3$ \\
\midrule
Feasible deterministic configurations (see (\ref{eq:feasible}))
	& $1\,056$
    & $160\,380$ \\
Meaningful deterministic configurations
	& 216
    & 43\,326 \\
Second stage configurations
	& 28
    & 4\,901 \\
Overall generation time, sec (wall time)
	& 293
    & 43\,326 \\
\bottomrule
\end{tabular}
\end{center}

\caption{CMCS generation data.}
\label{tab:cmcs-generation}
\end{table}

 The configurations generated for $\lambda = 2$ and $\lambda = 3$ are shown in Figure~\ref{fig:cmcs}.
 Each node in these diagrams corresponds to a component, and each arc to a transition.
 Blue arcs show transitions when the last component execution was successful (improved the solution) and red -- when unsuccessful (the solution quality was worsened or has not changed).
 The thickness of the arc indicates the frequency of that transition; it is proportional to square root of that frequency.

\tikzset{vertex/.style={circle, draw, thick}}
\tikzset{edge base/.style={->, >=stealth'}}
\tikzset{improved/.style={edge base, blue!#1!white, bend left=20}}
\tikzset{unimproved/.style={edge base, red!#1!white, bend left=20}}
\tikzset{loop improved/.style={edge base, blue!#1!white, loop above, in=70, out=110, looseness=6}}
\tikzset{loop unimproved/.style={edge base, red!#1!white, loop above, in=60, out=120, looseness=7}}

\begin{figure}[htb]
	\centering
    \begin{subfigure}[t]{0.38\textwidth}
    \footnotesize
	\begin{tikzpicture}
		\useasboundingbox (-2.3,-1.1) rectangle (2.3,2.3);
        
		\node[vertex, align=center] (RndClose4) at (1.4, 0.0) {Close Rnd\\ (4)};
		\node[vertex] (OpenBest) at (-1.4, 0.0) {Open Best};
		\path (RndClose4) edge[improved=100, line width=0.24074844657566, bend left=0] (OpenBest);
		\path (RndClose4) edge[unimproved=100, line width=3.53416751596637] (OpenBest);
		\path (OpenBest) edge[unimproved=100, line width=3.54235794993116] (RndClose4);
		\path (OpenBest) edge[loop improved=100, line width=5] (OpenBest);
   	\end{tikzpicture}
    
	\caption{Best performing two-component CMCS configuration}
    \label{fig:cmcs2}
	\end{subfigure}
    \hfill
    \begin{subfigure}[t]{0.55\textwidth}
    \footnotesize
	\begin{tikzpicture}
		\useasboundingbox (-3.2,-1.1) rectangle (3.2,2.3);

		\node[vertex, align=center] (RndOpen4) at (2.5, 0.0) {Open Rnd\\ (4)};
		\node[vertex] (CloseBest) at (-2.5, 0) {Close Best};
		\node[vertex, align=center] (RndBestExch) at (0, 0) {Exchange\\ Half Fixed};
		
        \path (RndOpen4) edge[unimproved=100, line width=2.5188202294687] (RndBestExch);
		\path (CloseBest) edge[loop improved=100, line width=5] (CloseBest);
		\path (CloseBest) edge[unimproved=100, line width=2.75246343406031] (RndBestExch);
		\path (RndBestExch) edge[unimproved=100, line width=2.5188202294687] (RndOpen4);
		\path (RndBestExch) edge[improved=100, line width=2.75246343406031] (CloseBest);
	\end{tikzpicture}
    
    \caption{Best performing three-component CMCS configuration}
    \label{fig:cmcs3}
    \end{subfigure}
    
	\caption{
    	Best performing CMCS configurations.
        The blue arcs correspond to successful transitions (after the solution was improved), and the red arcs correspond to the unsuccessful transitions (after the solution was not improved).
        The thickness of an arc indicates the frequency of the corresponding transition.}
    \label{fig:cmcs}
\end{figure}
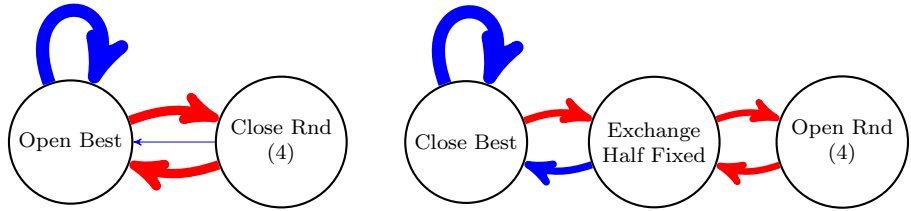

 The strategy of the two-component configuration is easy to explain.
 The configuration opens sites. in a greedy manner, as long as this improves the solution.
 Then it closes four sites randomly andm gets back to adding new sites.
 This strategy, in fact, exactly replicates the behaviour of iterated local search.

 The three-component configuration is more complex, but its logic can still be interpreted.
 The algorithm closes sites, in a greedy manner, until it reaches a local minimum.
 Then it attempts to replace some site with another one (`Exchange Half Fixed').
 If successful, there is a chance some other site got redundant which is why it returns to the `Close Best' component.
 If, however, `Exchange Half Fixed' fails, a mutation is applied.
 In particular, four random sites are opened and the control is passed back to `Exchange Half Fixed'.
 There is a high chance that `Exchange Half Fixed' will be able to improve the solution, and then the control will be returned to `Close Best'.
 Note that this strategy closely resembles variable neighbourhood search, with three neighbourhoods.
 As soon as the search in one neighbourhood fails, the next neighbourhood is used.
 It is non-typical though that one of the local searches (`Exchange Half Fixed') explores only a randomly chosen area of the neighbourhood.
 
 While we can explain the behaviour of the three-component CMCS configuration and even show its similarity to a well-known metaheuristic, the point of the automated generation is that this strategy was produced without prior knowledge of existing metaheuristics.
 Moreover, many decisions were taken automatically, such as which components to include in the metaheuristic, and in which order to use them.
 The whole design process is completely unbiased, hence the generated configuration is objectively one of the most effective ones possible within the framework.  
 (The exact definition of effectiveness may vary, but then the CMCS generation procedure can also be adjusted accordingly.)
 Thus, we can expect that the generated configuration performs at least as well as any human-designed metaheuristics, unless the limited component pool or complexity of CMCS is an issue.

 The source codes of our two- and three-component CMCS configurations can be downloaded from \\
 \url{http://csee.essex.ac.uk/staff/dkarap/splp-source-and-solutions.zip}.

\subsection{Experiments with the KG instances}
\label{sec:experiments}

 In this section we solve the KG instances with the two- and three-component CMCS configurations discussed above.
 We first solve the 50 instances to which optimal solutions are not yet known.
 We use time budget 7\,200 sec, same as in \cite{Fischetti2017}.  
 (The test machine used in~\cite{Fischetti2017} is based on Intel Xeon E3-1220V2 CPU (3.10 GHz), which is comparable to our CPUs.  
 However, Fischetti et al.~\cite{fischetti2017} utilised four CPU cores, effectively increasing computational power four-fold.
 By assuming that one time unit in the experiments of Fischetti et al.~\cite{fischetti2017} is equal to one time unit in our experiments, we give Fischetti et al.~\cite{fischetti2017} advantage.)
 
 Our results are reported in Table~\ref{tab:upper-bounds}.
 These new upper bounds are the best solutions we found in our experiments (as the reader will see later, all these solutions were obtained by the three-component CMCS configuration with time budget 1000 sec).
 We improved 12 best known upper bounds and matched all others.
 We further tested our solvers on the instances for which optimal solutions are known.
 Our three-component CMCS configuration could solve any of those instances to optimality within one second.
 While not being a formal proof, this suggests that we, perhaps, have also reached optimal solutions for most of the yet unsolved instances.
 
\begin{table}
\footnotesize
\begin{center}
\begin{tabular}{@{} lrrr @{}}
\toprule
Instance	&	Previously best known	&	Our best	&	Difference	\\
\midrule
ga500a-1	&	511\,383	&	511\,383	&	0	\\
ga500a-2	&	511\,255	&	511\,255	&	0	\\
ga500a-3	&	510\,810	&	510\,810	&	0	\\
ga500a-4	&	511\,008	&	511\,008	&	0	\\
ga500a-5	&	511\,239	&	511\,226	&	-13	\\
ga500b-1	&	538\,060	&	538\,060	&	0	\\
ga500b-2	&	537\,850	&	537\,850	&	0	\\
ga500b-3	&	537\,924	&	537\,921	&	-3	\\
ga500b-4	&	537\,925	&	537\,925	&	0	\\
ga500b-5	&	537\,482	&	537\,482	&	0	\\
ga750a-1	&	763\,528	&	763\,520	&	-8	\\
ga750a-2	&	763\,653	&	763\,623	&	-30	\\
ga750a-3	&	763\,697	&	763\,684	&	-13	\\
ga750a-4	&	763\,945	&	763\,941	&	-4	\\
ga750a-5	&	763\,786	&	763\,786	&	0	\\
ga750b-1	&	796\,454	&	796\,454	&	0	\\
ga750b-2	&	795\,963	&	795\,963	&	0	\\
ga750b-3	&	796\,130	&	796\,130	&	0	\\
ga750b-4	&	797\,013	&	797\,013	&	0	\\
ga750b-5	&	796\,387	&	796\,312	&	-75	\\
ga750c-1	&	902\,026	&	902\,026	&	0	\\
ga750c-2	&	899\,651	&	899\,651	&	0	\\
ga750c-3	&	900\,010	&	900\,010	&	0	\\
ga750c-4	&	900\,044	&	900\,044	&	0	\\
ga750c-5	&	899\,235	&	899\,235	&	0	\\
gs500a-1	&	511\,188	&	511\,187	&	-1	\\
gs500a-2	&	511\,179	&	511\,179	&	0	\\
gs500a-3	&	511\,112	&	511\,106	&	-6	\\
gs500a-4	&	511\,137	&	511\,137	&	0	\\
gs500a-5	&	511\,293	&	511\,293	&	0	\\
gs500b-1	&	537\,931	&	537\,931	&	0	\\
gs500b-2	&	537\,763	&	537\,763	&	0	\\
gs500b-3	&	537\,854	&	537\,854	&	0	\\
gs500b-4	&	537\,742	&	537\,742	&	0	\\
gs500b-5	&	538\,270	&	538\,270	&	0	\\
gs750a-1	&	763\,671	&	763\,671	&	0	\\
gs750a-2	&	763\,548	&	763\,548	&	0	\\
gs750a-3	&	763\,727	&	763\,702	&	-25	\\
gs750a-4	&	763\,887	&	763\,887	&	0	\\
gs750a-5	&	763\,614	&	763\,614	&	0	\\
gs750b-1	&	797\,026	&	797\,026	&	0	\\
gs750b-2	&	796\,170	&	796\,170	&	0	\\
gs750b-3	&	796\,589	&	796\,589	&	0	\\
gs750b-4	&	796\,734	&	796\,709	&	-25	\\
gs750b-5	&	796\,365	&	796\,365	&	0	\\
gs750c-1	&	900\,363	&	900\,363	&	0	\\
gs750c-2	&	897\,886	&	897\,886	&	0	\\
gs750c-3	&	901\,656	&	901\,089	&	-567	\\
gs750c-4	&	901\,239	&	901\,239	&	0	\\
gs750c-5	&	900\,216	&	900\,216	&	0	\\
\bottomrule
\end{tabular}
\end{center}

\caption{Previous and new upper bounds for the yet unsolved KG instances.}
\label{tab:upper-bounds}
\end{table}

 Our best solutions can be downloaded from\\ \url{http://csee.essex.ac.uk/staff/dkarap/splp-source-and-solutions.zip}, and are also reported in Appendices~\ref{sec:best-known-solutions-solved} and \ref{sec:best-known-solutions-unsolved}.

 In Table~\ref{tab:compare-to-fischetti}, we compare our two- and three-component CMCS configurations to the results of the previous attack on the KG instances Fischetti et al.~\cite{fischetti2017}.
 The time budget of each method is given in the second row of the table.
 Observe that either of the two CMCS configurations clearly outperforms~\cite{Fischetti2017} being given only 100 seconds, whereas the time budget in Fischetti et al.~\cite{fischetti2017} is 7200 seconds.
 Moreover, being given 1000 seconds, the three-component CMCS configuration matches or outperforms Fischetti et al.~\cite{fischetti2017} on every instance, finding all the new best solutions.
 Hence, the three-component CMCS is faster than Fischetti et al.~\cite{fischetti2017} by two orders of magnitude, and it is capable of achieving higher solution quality.

\begin{table}
\footnotesize
\begin{tabular}{@{} lr rrrrrrrrrrr @{}}
\toprule
Solver:	&	Fischetti et al.~\cite{fischetti2017} & \multicolumn{5}{l}{Two-component configuration} && \multicolumn{5}{l}{Three-component configuration}	\\
\cmidrule(lr){3-7}
\cmidrule(l){9-13}
Budget, sec: & 7200 & 1	&	10	&	100	&	1000	&	7200	&&	1	&	10	&	100	&	1000	&	7200	\\
\midrule
ga500a-1	&	511383	&	53	&	7	&	0	&	0	&	0	&&	0	&	0	&	0	&	0	&	0	\\
ga500a-2	&	511255	&	7	&	0	&	0	&	0	&	0	&&	21	&	0	&	0	&	0	&	0	\\
ga500a-3	&	510810	&	3	&	0	&	0	&	0	&	0	&&	7	&	0	&	0	&	0	&	0	\\
ga500a-4	&	511008	&	37	&	30	&	0	&	0	&	0	&&	48	&	30	&	30	&	0	&	0	\\
ga500a-5	&	511239	&	95	&	0	&	-13	&	-13	&	-13	&&	50	&	1	&	-13	&	-13	&	-13	\\
ga500b-1	&	538060	&	0	&	0	&	0	&	0	&	0	&&	0	&	0	&	0	&	0	&	0	\\
ga500b-2	&	537850	&	7	&	0	&	0	&	0	&	0	&&	0	&	0	&	0	&	0	&	0	\\
ga500b-3	&	537924	&	-3	&	-3	&	-3	&	-3	&	-3	&&	166	&	-3	&	-3	&	-3	&	-3	\\
ga500b-4	&	537925	&	43	&	0	&	0	&	0	&	0	&&	69	&	0	&	0	&	0	&	0	\\
ga500b-5	&	537482	&	0	&	0	&	0	&	0	&	0	&&	0	&	0	&	0	&	0	&	0	\\
ga750a-1	&	763528	&	141	&	0	&	0	&	-8	&	-8	&&	150	&	-8	&	18	&	-8	&	-8	\\
ga750a-2	&	763653	&	63	&	8	&	-30	&	-30	&	-30	&&	59	&	-19	&	-17	&	-30	&	-30	\\
ga750a-3	&	763697	&	196	&	46	&	-13	&	-13	&	-13	&&	170	&	59	&	-7	&	-13	&	-13	\\
ga750a-4	&	763945	&	200	&	34	&	-4	&	19	&	-4	&&	180	&	103	&	-4	&	-4	&	-4	\\
ga750a-5	&	763786	&	260	&	12	&	4	&	8	&	0	&&	212	&	63	&	0	&	0	&	0	\\
ga750b-1	&	796454	&	337	&	0	&	0	&	0	&	0	&&	26	&	0	&	0	&	0	&	0	\\
ga750b-2	&	795963	&	190	&	0	&	0	&	0	&	0	&&	248	&	0	&	0	&	0	&	0	\\
ga750b-3	&	796359	&	132	&	-216	&	-229	&	-229	&	-229	&&	90	&	-229	&	-229	&	-229	&	-229	\\
ga750b-4	&	797013	&	128	&	0	&	0	&	0	&	0	&&	0	&	0	&	0	&	0	&	0	\\
ga750b-5	&	796549	&	-104	&	-237	&	-237	&	-237	&	-237	&&	-65	&	-214	&	-237	&	-237	&	-237	\\
ga750c-1	&	902026	&	0	&	0	&	0	&	0	&	0	&&	0	&	0	&	0	&	0	&	0	\\
ga750c-2	&	899651	&	81	&	0	&	0	&	0	&	0	&&	81	&	0	&	0	&	0	&	0	\\
ga750c-3	&	900019	&	-9	&	-9	&	-9	&	-9	&	-9	&&	-9	&	-9	&	-9	&	-9	&	-9	\\
ga750c-4	&	900044	&	0	&	0	&	0	&	0	&	0	&&	0	&	0	&	0	&	0	&	0	\\
ga750c-5	&	899235	&	0	&	0	&	0	&	0	&	0	&&	0	&	0	&	0	&	0	&	0	\\
gs500a-1	&	511188	&	12	&	-1	&	-1	&	-1	&	-1	&&	114	&	41	&	-1	&	-1	&	-1	\\
gs500a-2	&	511179	&	0	&	0	&	0	&	0	&	0	&&	0	&	0	&	0	&	0	&	0	\\
gs500a-3	&	511112	&	0	&	25	&	-6	&	-6	&	-6	&&	31	&	0	&	-6	&	-6	&	-6	\\
gs500a-4	&	511137	&	117	&	0	&	0	&	0	&	0	&&	139	&	0	&	0	&	0	&	0	\\
gs500a-5	&	511293	&	81	&	27	&	27	&	0	&	0	&&	88	&	0	&	0	&	0	&	0	\\
gs500b-1	&	537931	&	64	&	0	&	0	&	0	&	0	&&	0	&	0	&	0	&	0	&	0	\\
gs500b-2	&	537763	&	16	&	0	&	0	&	0	&	0	&&	48	&	0	&	0	&	0	&	0	\\
gs500b-3	&	537854	&	72	&	0	&	0	&	0	&	0	&&	0	&	0	&	0	&	0	&	0	\\
gs500b-4	&	537742	&	0	&	0	&	0	&	0	&	0	&&	0	&	0	&	0	&	0	&	0	\\
gs500b-5	&	538270	&	82	&	0	&	0	&	0	&	0	&&	82	&	0	&	0	&	0	&	0	\\
gs750a-1	&	763671	&	63	&	5	&	0	&	0	&	0	&&	159	&	17	&	0	&	0	&	0	\\
gs750a-2	&	763548	&	199	&	15	&	15	&	0	&	0	&&	157	&	15	&	14	&	0	&	0	\\
gs750a-3	&	763727	&	155	&	46	&	21	&	-25	&	-25	&&	-12	&	11	&	21	&	-25	&	-25	\\
gs750a-4	&	763922	&	58	&	6	&	-35	&	-35	&	-35	&&	53	&	22	&	-3	&	-8	&	-35	\\
gs750a-5	&	763614	&	102	&	27	&	2	&	0	&	0	&&	87	&	18	&	2	&	0	&	0	\\
gs750b-1	&	797329	&	-138	&	-303	&	-303	&	-303	&	-303	&&	-303	&	-303	&	-303	&	-303	&	-303	\\
gs750b-2	&	796170	&	31	&	25	&	0	&	0	&	0	&&	31	&	31	&	0	&	0	&	0	\\
gs750b-3	&	796589	&	0	&	0	&	0	&	0	&	0	&&	534	&	0	&	0	&	0	&	0	\\
gs750b-4	&	797020	&	-311	&	-286	&	-311	&	-311	&	-311	&&	-178	&	-286	&	-311	&	-311	&	-311	\\
gs750b-5	&	796365	&	0	&	0	&	0	&	0	&	0	&&	0	&	0	&	0	&	0	&	0	\\
gs750c-1	&	900363	&	0	&	0	&	0	&	0	&	0	&&	0	&	0	&	0	&	0	&	0	\\
gs750c-2	&	897886	&	0	&	0	&	0	&	0	&	0	&&	187	&	0	&	0	&	0	&	0	\\
gs750c-3	&	901656	&	-567	&	-567	&	-567	&	-567	&	-567	&&	-567	&	-567	&	-567	&	-567	&	-567	\\
gs750c-4	&	901239	&	0	&	0	&	0	&	0	&	0	&&	0	&	0	&	0	&	0	&	0	\\
gs750c-5	&	900216	&	0	&	0	&	0	&	0	&	0	&&	0	&	0	&	0	&	0	&	0	\\
\midrule			
\multicolumn{2}{r}{Improved}	&	6	&	8	&	14	&	15	&	16	&&	6	&	9	&	14	&	16	&	16	\\
\multicolumn{2}{r}{Same}	&	14	&	28	&	31	&	33	&	34	&&	16	&	29	&	31	&	34	&	34	\\
\multicolumn{2}{r}{Worse}	&	30	&	14	&	5	&	2	&	0	&&	28	&	12	&	5	&	0	&	0	\\
    
\bottomrule
\end{tabular}

\caption{
	Comparison of the CMCS configurations to~Fischetti et al.~\cite{fischetti2017}.}
\label{tab:compare-to-fischetti}
\end{table}

 We also note here that the three-component CMCS configuration performs better, on average, than the two-component one.
 For example, the three-component CMCS configuration given 1000 seconds achieves the same solution quality as the two-component CMCS configuration given 7200 seconds.
 The two-component configuration was also less successful on the solved KG instances; even given 7\,200 seconds per instances, it could not reach the optimal solution for one of them.
 This demonstrates the importance of configuration complexity and diversity of components; setting $\Lambda = 2$ could be considered as a minimal option, which restricts the performance that can be achieved by corresponding configurations.
 On the other hand, we have evidence that $\Lambda = 3$ is sufficient to achieve outstanding performance when compared to human-designed algorithms.
 A more efficient CMCS generation routine will let us verify if a four-component configuration can achieve an even better performance.

\section{Conclusions}
\label{sec:conclusions}

 In this paper, we discussed automated generation of CMCS configurations for the SPLP, and have shown the success of our approach.
 In particular, we clearly outperformed the previous state-of-the-art solver and improved the best known upper bounds for 12 out of 50 yet unsolved KG instances.
 The outstanding performance of our SPLP heuristic is attributed to that it was generated automatically.
 
 The automated generation of the algorithm has several obvious advantages.
 One is that it saves labour and human expertise required for (meta)heuristic design.
 Also, automated generation is significantly quicker than a manual design process, hence the entire algorithm design can be completed within a few days.
 Finally, the computer is capable of testing more combinations than a human and objectively selecting the best of them.
 This lack of bias means, among other things, that the computer does test strategies that a human would usually rule out, and in our experience such unusual strategies often demonstrate unexpectedly good performance.
 
 In the spirit of the no free lunch theorem, we note here that the selected configuration performs best only under certain circumstances such as a specific instance family or certain time budget.
 By correctly selecting the training dataset, we can obtain an algorithm that is best suited for our particular case.
 Moreover, it is easy to obtain several algorithms for various circumstances and requirements and then use the most appropriate one for each job.
 
 In this project, we limited CMCS configurations to deterministic strategies, and also restricted the number of components to be included in a configuration.
 These simple measures greatly reduced the number of candidate configurations allowing us to enumerate all of them and choose the best performing one.
 We use a combination of Pareto dominance and sign test to quickly rule out less promising configurations, and then apply a multi-criteria optimisation method to choose a single best candidate.
 
 We leave for future work investigation of more efficient CMCS generation procedures, which will allow one to include more components into the component pool and consider more sophisticated configurations.
 Also, it would be interesting to select not a single best configuration but several well-performing configurations with complementary properties, and select the most appropriate one at run time.
 Finally, we are interested to apply the CMCS approach to other classes of allocation and cliustering instances as well as to routing and scheduling optimisation problems.

\bibliographystyle{plain}
\bibliography{refs}

\appendix
\section{Optimal solutions for instances solved to optimality}
\label{sec:best-known-solutions-solved}

\begin{longtable}{@{}lrp{0.7\linewidth}@{}} 
\toprule
Instance & Obj.\ v. & Opened sites \\
\midrule
\endhead

\bottomrule
\endfoot

ga250a-3	&257953	&22 35 39 46 57 66 76 86 97 100 105 112 114 116 121 124 126 127 144 154 155 176 192 196 200 207 211 219 223 227 229 237 246 249	\\
ga250a-5	&258190	&13 17 29 35 40 43 49 55 60 63 79 82 110 126 135 139 150 157 161 174 178 179 198 201 204 208 211 230 232 241 248	\\
ga500c-5	&621313	&4 75 183 259 360 491	\\
gs500c-3	&621204	&98 195 216 245 333 429	\\
gs500c-5	&623180	&22 51 276 355 439 444	\\
ga250a-1	&257957	&21 32 38 47 53 56 58 84 94 100 101 103 111 129 136 139 144 146 149 150 168 170 175 178 203 204 219 224 234 238 239 250	\\
ga250a-2	&257502	&11 37 55 62 64 84 88 99 100 103 107 114 115 116 118 132 146 157 158 160 171 191 200 211 213 217 218 221 237 238 240 250	\\
ga250a-4	&257987	&4 5 7 30 31 37 53 55 69 73 74 75 84 92 93 103 108 115 119 127 129 153 163 168 173 174 188 199 200 203 208 213 219 236 250	\\
ga250b-1	&276296	&10 56 60 94 106 129 149 150 170 203 219 250	\\
ga250b-2	&275141	&37 55 88 103 135 141 158 191 211 213 231	\\
ga250b-3	&276093	&1 18 22 35 39 50 97 192 200 229 246	\\
ga250b-4	&276332	&5 7 36 56 77 92 124 160 228 236 250	\\
ga250b-5	&276404	&40 57 79 110 157 161 183 184 208 211 241 246	\\
ga250c-1	&334135	&100 154 175 231	\\
ga250c-2	&330728	&45 55 88 99	\\
ga250c-3	&333662	&22 97 127 138	\\
ga250c-4	&332423	&5 124 143 188	\\
ga250c-5	&333538	&74 110 157 247	\\
ga500c-1	&621360	&127 269 378 403 430	\\
ga500c-2	&621464	&28 107 212 315 344 456	\\
ga500c-3	&621428	&68 187 314 326 370 474	\\
ga500c-4	&621754	&56 97 307 350 436	\\
gs250a-1	&257964	&4 10 12 25 27 30 47 51 63 71 119 123 126 132 137 143 145 155 161 163 169 176 177 178 203 214 232 234 236 238 245 246 249	\\
gs250a-2	&257573	&9 24 25 40 43 46 52 74 77 86 87 88 95 96 98 100 101 113 114 120 130 139 154 160 161 165 166 184 191 241 245 250	\\
gs250a-3	&257626	&9 20 33 34 37 38 55 60 67 69 71 72 91 110 120 121 132 139 144 148 166 172 174 177 187 189 190 199 204 209 223 229 234	\\
gs250a-4	&257961	&3 20 31 36 46 54 101 102 104 115 118 128 139 143 144 159 160 163 168 179 188 193 195 207 208 217 221 226 233 237	\\
gs250a-5	&257896	&18 33 36 47 49 60 76 77 89 98 104 114 118 122 124 133 137 156 161 168 172 189 204 207 209 212 213 217 223 227 228 230 235 250	\\
gs250b-1	&276761	&8 27 47 63 71 113 137 145 170 178 229 232	\\
gs250b-2	&275675	&25 43 52 69 77 87 120 139 149 160 221 245	\\
gs250b-3	&275710	&32 55 57 60 67 69 82 166 172 174 210	\\
gs250b-4	&276114	&13 19 31 35 97 106 139 144 157 177 191 247	\\
gs250b-5	&275916	&18 36 118 122 124 137 166 172 177 204 209 230	\\
gs250c-1	&332935	&63 170 176 232	\\
gs250c-2	&334630	&25 52 83 144	\\
gs250c-3	&333000	&57 60 69 166	\\
gs250c-4	&333158	&52 144 157 191	\\
gs250c-5	&334635	&18 84 114 186	\\
gs500c-1	&620041	&29 102 112 242 440	\\
gs500c-2	&620434	&70 286 424 439 495	\\
gs500c-4	&620437	&96 247 283 316 390	\\
\end{longtable}


\section{Best known solutions for instances not yet solved to optimality}
\label{sec:best-known-solutions-unsolved}

 The objective values of these solutions can be found in Table~\ref{tab:upper-bounds}, column `Our best'.

\begin{longtable}{@{}lp{0.85\linewidth}@{}}
\toprule
Instance & Opened sites \\
\midrule
\endhead

\bottomrule
\endfoot

ga500a-1	&22 28 52 59 65 70 73 79 86 90 100 103 111 126 142 152 156 173 177 189 199 205 208 219 221 234 245 246 260 265 269 275 280 299 301 303 313 375 378 397 410 419 430 463 475 486 490 494	\\
ga500a-2	&34 51 54 70 84 116 120 122 126 144 155 158 169 177 188 193 195 204 212 213 218 222 223 238 255 289 313 315 321 333 338 345 360 372 391 397 399 401 413 415 437 450 466 470 476 478 485 487 500	\\
ga500a-3	&12 33 36 37 44 49 55 65 75 85 88 92 94 96 114 132 145 148 150 166 178 184 185 187 192 196 201 220 241 252 257 278 285 288 290 303 340 364 367 370 375 378 386 393 431 451 474	\\
ga500a-4	&18 19 29 35 39 42 43 49 56 65 74 78 102 119 138 140 144 155 188 197 204 214 267 273 280 281 282 293 317 329 340 346 350 360 364 371 377 388 404 411 417 419 430 436 448 456 466 484 496	\\
ga500a-5	&4 11 14 22 34 36 38 40 47 51 55 95 100 120 123 125 127 133 155 174 181 183 199 216 229 283 284 301 316 321 326 328 332 336 348 369 371 380 382 387 390 397 399 424 429 487 488 491 497	\\
ga500b-1	&34 100 127 153 156 176 184 189 199 236 277 375 378 379 410 430 470	\\
ga500b-2	&28 34 51 137 212 213 225 238 241 245 249 268 315 336 338 344 459 478	\\
ga500b-3	&36 66 92 94 166 185 187 189 241 290 300 303 326 340 370 393 431 474	\\
ga500b-4	&24 138 204 282 293 329 330 343 360 388 396 436 448 451 456 466 484 496	\\
ga500b-5	&2 14 55 123 124 135 142 147 181 183 231 258 259 349 360 382 399 414	\\
gs500a-1	&6 22 42 53 55 58 94 95 115 116 120 121 126 127 129 144 149 154 164 171 173 212 239 252 270 285 294 300 320 321 327 335 336 343 352 377 379 384 385 389 399 420 426 429 434 442 464 490	\\
gs500a-2	&9 50 53 62 70 99 109 110 115 124 168 169 175 185 198 202 204 218 229 233 241 247 269 276 289 290 294 295 301 316 333 335 336 356 358 376 383 394 400 422 426 437 439 453 457 459 460 463 464 470	\\
gs500a-3	&7 17 28 38 41 55 65 67 74 84 86 110 117 147 152 153 162 173 212 219 223 244 256 259 269 271 273 287 300 301 308 310 365 369 371 377 381 385 394 401 413 417 421 437 453 456 493 494	\\
gs500a-4	&9 10 14 18 56 67 68 84 87 93 95 123 124 136 137 161 165 173 180 189 194 196 202 217 229 231 258 273 277 281 290 350 356 359 363 371 378 380 390 391 435 438 453 458 464 484 490 491 495	\\
gs500a-5	&3 4 7 13 15 30 40 47 60 69 78 86 122 123 136 153 156 159 160 171 174 185 192 231 233 234 235 250 251 257 268 281 304 312 316 322 331 338 384 391 411 424 431 444 459 460 481 498	\\
gs500b-1	&22 29 45 82 102 112 116 193 215 257 258 313 385 410 440 468 470 471	\\
gs500b-2	&24 70 85 95 115 168 233 247 329 356 358 382 383 408 437 439 457 488	\\
gs500b-3	&7 41 116 216 235 245 255 269 273 279 287 299 308 333 347 371 392 429	\\
gs500b-4	&76 91 103 120 132 142 165 171 173 212 230 351 380 406 437 438 447 491	\\
gs500b-5	&3 7 13 40 65 105 153 185 226 235 319 322 372 384 431 444 460 486	\\
ga750a-1	&2 18 35 50 52 67 69 71 74 105 110 111 117 118 127 152 155 209 219 233 234 242 263 280 296 305 309 316 330 335 346 381 430 431 435 446 449 457 478 484 494 512 538 540 548 559 564 587 616 640 644 647 650 667 680 689 711 713 716 729 738 745	\\
ga750a-2	&1 18 40 45 71 102 104 109 110 114 120 126 131 144 150 154 168 170 180 183 211 214 227 234 235 237 239 259 283 288 289 296 301 316 351 352 357 359 362 369 375 382 387 436 447 455 491 511 528 530 539 550 581 582 615 642 644 645 673 684 710 722	\\
ga750a-3	&49 68 71 75 88 95 101 109 113 127 183 202 206 212 254 266 295 298 334 339 349 356 361 379 389 400 415 419 428 433 436 439 446 447 450 464 465 483 506 560 561 565 570 575 580 581 585 590 606 626 627 645 657 666 669 679 682 694 698 731	\\
ga750a-4	&25 54 79 87 99 100 104 108 112 149 154 160 163 169 176 195 224 226 258 266 271 279 291 292 303 305 319 324 326 363 386 400 413 416 418 420 437 454 468 487 496 534 536 551 561 568 628 635 636 652 656 669 672 684 693 695 703 718 733 737 748	\\
ga750a-5	&18 34 35 67 75 77 80 87 93 117 119 146 161 167 168 187 189 195 224 228 235 246 247 271 315 320 325 329 359 365 367 373 389 411 421 424 426 429 452 456 475 476 507 522 523 524 532 553 562 565 578 588 655 678 702 706 709 713 734 736 747 749	\\
ga750b-1	&58 67 71 100 117 184 214 335 346 386 478 484 512 559 589 593 616 647 662 711 720 745 746	\\
ga750b-2	&1 45 109 110 144 168 182 214 235 237 239 283 288 296 308 329 375 505 637 644 645 712	\\
ga750b-3	&53 68 101 202 206 215 254 298 334 356 404 408 464 560 570 575 596 604 669 673 726 728	\\
ga750b-4	&47 52 55 104 115 128 149 154 202 232 266 303 358 434 468 551 635 639 672 704 733	\\
ga750b-5	&29 42 49 87 99 182 193 194 218 224 351 363 373 376 380 456 473 523 667 685 700 746	\\
ga750c-1	&214 418 476 587 593 644 711	\\
ga750c-2	&1 170 182 235 237 564 590	\\
ga750c-3	&68 101 173 215 439 548 616	\\
ga750c-4	&128 144 154 279 413 456 704	\\
ga750c-5	&14 344 376 456 577 659 685	\\
gs750a-1	&2 20 52 64 75 89 93 124 128 130 143 147 177 199 205 218 232 236 237 240 264 279 281 289 305 317 320 335 336 374 393 403 404 425 453 458 466 468 482 487 496 518 538 541 545 554 555 564 602 609 637 651 658 659 663 669 672 681 682 706 713 729 743	\\
gs750a-2	&1 19 24 62 64 70 92 93 94 100 108 113 134 137 146 157 178 180 186 213 225 265 268 278 281 325 334 336 341 348 362 393 397 398 411 444 451 460 462 493 494 498 504 554 574 575 594 595 620 625 628 636 639 646 650 661 721 724 735	\\
gs750a-3	&6 22 35 38 46 66 96 110 117 119 133 135 171 182 192 230 250 287 288 291 317 354 357 360 367 374 385 392 396 409 412 425 437 450 458 463 465 487 499 510 528 554 561 562 583 596 609 612 624 640 677 688 702 710 715 721 728 732 746	\\
gs750a-4	&2 4 24 26 34 56 78 80 81 95 107 115 117 123 136 139 145 146 151 172 174 190 241 243 258 266 269 294 305 323 332 346 377 388 399 412 434 443 459 473 477 498 500 522 530 536 537 551 558 573 603 617 640 641 656 666 669 673 721 722 740 749	\\
gs750a-5	&3 12 31 46 54 65 74 88 91 96 102 104 112 114 119 135 150 182 198 202 207 231 234 288 302 317 356 359 386 394 397 410 411 417 421 425 483 579 582 587 607 612 638 644 654 662 663 669 672 680 693 707 716 720 721 725 731 744	\\
gs750b-1	&41 45 67 75 128 130 213 232 237 242 279 281 313 428 487 538 573 658 698 699 725 743	\\
gs750b-2	&1 10 108 126 191 213 214 257 265 314 336 341 348 367 444 450 494 508 617 636 639 650 661	\\
gs750b-3	&22 26 64 135 171 192 281 291 304 317 446 462 484 553 561 583 702 706 715 727 728	\\
gs750b-4	&4 59 81 95 107 113 165 174 179 190 238 248 261 294 305 355 431 459 616 641 654 721	\\
gs750b-5	&3 31 74 90 104 200 281 302 359 392 394 427 487 540 549 568 593 627 635 693 707 709	\\
gs750c-1	&67 112 128 428 479 603 639	\\
gs750c-2	&29 92 108 265 336 494 639	\\
gs750c-3	&6 44 304 583 624 680 698	\\
gs750c-4	&26 139 151 287 522 628 721	\\
gs750c-5	&104 198 302 557 607 635 707	\\
\end{longtable}


\end{document}